%% file: manuscript.tex
\documentclass[journal,twoside,web]{ieeecolor}
\usepackage{etoolbox}
\makeatletter
\@ifundefined{color@begingroup}%
{\let\color@begingroup\relax
\let\color@endgroup\relax}{}%
\def\fix@ieeecolor@hbox#1{%
\hbox{\color@begingroup#1\color@endgroup}}
\patchcmd\@makecaption{\hbox}{\fix@ieeecolor@hbox}{}{\FAILED}
\patchcmd\@makecaption{\hbox}{\fix@ieeecolor@hbox}{}{\FAILED}

\usepackage{tmi}
\usepackage{cite}
\usepackage{amsmath,amssymb,amsfonts}
\usepackage{algorithmic}
\usepackage{graphicx}
\usepackage{textcomp}
\usepackage[ruled,linesnumbered]{algorithm2e}
\usepackage{verbatim}
\usepackage{multirow}
\usepackage{amssymb}
\usepackage{amsmath}
\usepackage{booktabs}
\usepackage{url}
\usepackage{tablefootnote}
\usepackage{stfloats}
\usepackage{color}
\usepackage{verbatim}
\usepackage{bigstrut}
\usepackage{makecell}
\usepackage{caption}
\usepackage{hyperref}
\usepackage{float}

\input{math_commands.tex}

\def\BibTeX{{\rm B\kern-.05em{\sc i\kern-.025em b}\kern-.08em
    T\kern-.1667em\lower.7ex\hbox{E}\kern-.125emX}}
\begin{document}
\title{
  BrainMass: Advancing Brain Network Analysis for Diagnosis with Large-scale Self-Supervised Learning
  
  }
\author{Yanwu Yang, \IEEEmembership{Member, IEEE}, Chenfei Ye, Guinan Su, Ziyao Zhang, Zhikai Chang,\\ Hairui Chen,  Piu Chan,  Yue Yu, Ting Ma \IEEEmembership{Member, IEEE}
\thanks{This work was done by Yanwu Yang during his internship at Peng Cheng Laboratory. The study is supported by grants from the National Natural Science Foundation of China (62276081), The National Key Research and Development Program of China (2021YFC2501202), and the Major Key Project of PCL. Corresponding author: Yue Yu, Ting Ma.}
\thanks{Yanwu Yang and Chenfei Ye contributed equally to this work.}
\thanks{Yanwu Yang, Hairui Chen, and Ting Ma are with the School of Electronics and Information Engineering, Harbin Institute of Technology at Shenzhen, Shenzhen, China, and the Peng Cheng Laboratory, Shenzhen, Guangdong, China. (e-mail: 20b952019@stu.hit.edu.cn, 23b952017@stu.hit.edu.cn, tma@hit.edu.cn)}
\thanks{Piu Chan is with Xuanwu Hospital, Capital Medical University, Beijing, China. (E-mail: pbchan@hotmail.com)}
\thanks{Guinan Su is with the Tencent Data Platform, Shenzhen 518057, China. (E-mail: guinansu33@gmail.com)}
\thanks{Ziyao Zhang is with the Paul C. Lauterbur Research Center for Biomedical Imaging, Shenzhen Institutes of Advanced Technology, Chinese Academy of Sciences, Shenzhen, Guangdong 518000, China and Peng Cheng Laboratory, Shenzhen, Guangdong 518000, China. (e-mail: zhangzy@pcl.ac.cn)}
\thanks{Yue Yu is with the Peng Cheng Laboratory, Shenzhen, Guangdong, China. (e-mail: yuy@pcl.ac.cn)}
\thanks{Chenfei Ye and Zhikai Chang are with the Harbin Institute of Technology at Shenzhen, Shenzhen, China (e-mail: Chenfei.ye@foxmail.com, 22b352010@stu.hit.edu.cn)}
}

\maketitle

\begin{abstract}


Foundation models pretrained on large-scale datasets via self-supervised learning demonstrate exceptional versatility across various tasks. Due to the heterogeneity and hard-to-collect medical data, this approach is especially beneficial for medical image analysis and neuroscience research, as it streamlines broad downstream tasks without the need for numerous costly annotations. However, there has been limited investigation into brain network foundation models, limiting their adaptability and generalizability for broad neuroscience studies. In this study, we aim to bridge this gap. In particular, (1) we curated a comprehensive dataset by collating images from 30 datasets, which comprises 70,781 samples of 46,686 participants. Moreover, we introduce pseudo-functional connectivity (pFC) to further generates millions of augmented brain networks by randomly dropping certain timepoints of the BOLD signal. (2) We propose the BrainMass framework for brain network self-supervised learning via mask modeling and feature alignment. BrainMass employs Mask-ROI Modeling (MRM) to bolster intra-network dependencies and regional specificity. Furthermore, Latent Representation Alignment (LRA) module is utilized to regularize augmented brain networks of the same participant with similar topological properties to yield similar latent representations by aligning their latent embeddings. Extensive experiments on eight internal tasks and seven external brain disorder diagnosis tasks show BrainMass's superior performance, highlighting its significant generalizability and adaptability. Nonetheless, BrainMass demonstrates powerful few/zero-shot learning abilities and exhibits meaningful interpretation to various diseases, showcasing its potential use for clinical applications.

\end{abstract}

\begin{IEEEkeywords}
Self-supervised learning, brain network, Transformer, large-scale, pretrain
\end{IEEEkeywords}

\section{Introduction}

Functional Magnetic Resonance Imaging (fMRI) utilizing the blood-oxygen-level-dependent (BOLD) effect has become an instrumental tool in neuroscience, offering a unique opportunity to in-vivo map the neural substrates of cognition \cite{logothetis2008we,heeger2002does,logothetis2001neurophysiological}. A key outcome of this paradigm is the development of functional brain networks, which are established through correlations between BOLD signal from various regions of interest (ROIs) to estimate the neural interactions and temporal synchronization \cite{fingelkurts2005functional}. These networks have become indispensable tools for brain disorder analysis, examining the underlying disconnectome in various diseases \cite{yang2021alteration,bastos2016tutorial}.

In recent years, the field of brain functional network analysis has been greatly influenced by deep learning approaches, which characterize complex interactions of ROIs with non-linear and deep embedded representations, and significantly improves the disease diagnosis performance. These include a range of techniques such as convolutional neural networks (CNN) \cite{kawahara2017brainnetcnn,huang2017modeling,huang2020identifying}, graph neural networks (GNN) \cite{zhao2022dynamic, li2021braingnn, yang2023mapping}, and Transformer networks \cite{kan2022brain,zhu2022multimodal}. Despite significant progress, a pervasive limitation of these studies is their limited generalizability and adaptability \cite{wen2023graph,jung2021inter}. 
Task-specific models are still the main methods used, that are limited by the number of annotated samples and poor adaptation to other tasks. And the lack of capabilities for few-shot or zero-shot learning, limits their potential use in clinical scenarios where only a few MRIs with annotations are available. Moreover, the data heterogeneity also hampers the generalizability \cite{segal2023regional,zhang2019integrating}.

One way to address this issue is through large-scale self-supervised learning (SSL) to produce homogeneous and generalizable representations. This method has shown promise, leading to impressive performance gains in a wide variety of downstream tasks across other domains \cite{he2020momentum,chen2020improved,devlin2018bert,conneau2019cross}. Unlike traditional pretrained models, foundation models pre-trained on large-scale datasets can handle a wide variety of tasks with a single set of model weights \cite{zhang2023challenges}.
However, in the field of medical image analysis, developing foundation models, in particular for brain networks, presents a significant challenge due to the limited data samples and insufficient self-supervised learning. Current studies leveraging SSL for brain network only achieve comparable performance to non-SSL methods \cite{jung2021inter,wen2023graph,hu2023brainnpt}. 
Consequently, specific foundation models on brain network is urgently needed in this field at the moment.


To this end, we aim to bridge the gap in foundation models for brain networks. In this paper, we curated a large cohort comprising 70,781 samples of 46,686 participants in multiple centers. We also introduce an augmentation method to create more brain networks that involves randomly dropping timepoints in the BOLD signals to pseudo-functional connectivity (pFC).
Moreover, we propose BrainMass, the first foundation model specifically designed for \textbf{Brain} network analysis with \textbf{M}ask modeling and representation \textbf{A}lignment via \textbf{S}elf-\textbf{S}upervised learning to pre-train the Transformer encoder:

(1) \textbf{MRM}: MRM is executed by randomly masking some ROIs and predicting the masked features by the remaining. In particular, classification heads are utilized to predict the meta labels (indices of the masked ROIs), and reconstruction heads are deployed to estimate the features of the masked ROIs. This inclusion helps relate intra-network dependencies and enhances locality characteristics for downstream tasks. 

(2) \textbf{LRA}: BrainMass leveraging LRA employs a dual-branch approach to extract representations from two pFCs derived from the same BOLD signal and regularizes them to achieve similar latent embeddings. 
This design acknowledges that augmented brain networks derived from the same participant should yield similar latent representations.
We leverage a dual branch network to extract the embeddings of two pFCs and regularize the them to be closer. 

To evaluate the effectiveness of our BrainMass, eight internal and seven external diagnosis tasks were carried out as the downstream tasks. We extracted the learned representations from BrainMass and employed an SVM classifier for classification. Our extensive experimental results demonstrated that BrainMass not only outperformed existing models but also exhibited remarkable generalizability, adaptability, and few-shot capabilities. Our key contributions are outlined as follows:

\begin{itemize}
  \item We built a large cohort comprising 46,686 participants with 70,781 samples for large-scale brain network studies. Furthermore, we developed augmented brain networks using pseudo-functional connectivity (pFC) to further enlarge the training set.

  \item We introduced the first brain network foundation model, BrainMass, and demonstrated its superior diagnostic performance, along with its impressive generalizability and adaptability across eight internal and seven external tasks.
  
  \item  Our explanatory analysis revealed that BrainMass is capable of identifying the patterns of various brain disorders and pinpointing meaningful key biomarkers.
  
  \item Our BrainMass exhibits powerful capabilities in zero-shot and few-shot learning, showcasing its potential for clinical applications.
  
  
\end{itemize} 

Our project is publicly online at \url{https://github.com/podismine/BrainMass}. The pre-trained weights are available for researchers to easily adapt the model for various tasks and analyze the biomarkers without the need for computationally expensive supervised fine-tuning.

\section{Related Works}\label{related works}
\subsection{Brain network study}
Significant advancements have been made over the past decade in the application of neuroimaging techniques to uncover alterations in brain network associated with various brain disorders.
Convolutional neural networks (CNN) are firstly proposed to facilitate end-to-end disease identification with promising performances and have been widely applied for analyzing network patterns such as BrainNetCNN \cite{kawahara2017brainnetcnn} and Deep Convolutional Auto-Encoder \cite{huang2017modeling}. 
A weighted correlation kernel-based convolutional neural network is built for learning the hierarchical features \cite{jie2020designing}.
In addition to CNNs, graph neural networks (GNNs) have gained prominence. GNNs have the capacity to capture information about neighboring structures within the brain. BrainGNN, for instance, introduced ROI-aware graph convolutional layers and ROI-selection pooling layers to predict neurological biomarkers at both the group and individual levels \cite{li2021braingnn}. Another approach, proposed by \cite{ktena2018metric}, involved learning a graph similarity metric using a siamese graph convolutional neural network. 
A dynamic graph network is proposed by learning from sparse connections among brain regions calculated dynamically from graph features \cite{zhao2022dynamic}. 
M-GCN regularized convolution on functional connectivity with structural graph Laplacian \cite{dsouza2021m}. Cross-GNN is proposed to capture inter-modal dependencies \cite{yang2023mappinga}.
Another type of GNN build transductive graphs and implement semi-supervised learning to predict via a node classification task \cite{parisot2018disease,song2021graph,yang2023deep}. 


More recently, the Transformer architecture has garnered considerable attention due to its exceptional performance in graph representation learning. However, most existing Transformer-based networks \cite{ying2021transformers,kreuzer2021rethinking} have achieved only limited success in brain network analysis. To address this limitation, BrainNetTransformer \cite{kan2022brain} was introduced with distinguishable cluster-aware embeddings to determine similar behaviors, which outperforms most of existing studies. 
However, these task-specific models are limited by the number of annotated samples, which limits their adaptation and generalization abilities to other tasks.

\subsection{Self-supervised Learning}
Self-supervised learning paradigms have delivered promising results in computer vision \cite{he2020momentum,chen2020improved} and natural language processing \cite{devlin2018bert,conneau2019cross}. These paradigms have introduced pre-trained foundation models that leverage self-supervised learning on extensive unannotated data. This approach produces standardized and generalized representations, offering substantial benefits across domains with limited task-specific data availability.
However, this paradigm has been investigated by only a few studies for brain networks. BrainNPT \cite{hu2023brainnpt}, for example, constructs disturbance inputs by replacing regional features to enhance the models' understanding of the underlying input patterns. 
\cite{jung2021inter} leverages a masked seed-based strategy for pretraining. BrainGSLs \cite{wen2023graph}, similarly, proposes an ensemble masked graph self-supervised framework based on masking and prediction. Nevertheless, these studies have only achieved modest improvements when compared to approaches without pre-training (i.e., approximately 71.5\% accuracy on the ABIDE dataset). It's important to note that these pre-training strategies, borrowed from BERT-like models, still rely on a substantial amount of training data to establish data dependencies. Nonetheless, graph neural networks are limited by the depth of the models to be applied for the foundation model studies, due to the oversmoothing issue \cite{chen2020measuring}. 
In summary, there remains a notable gap in the development of self-supervised learning studies tailored to uncover the intrinsic characteristics of brain network.
\begin{figure*}
  \centering
  \includegraphics[scale=0.9]{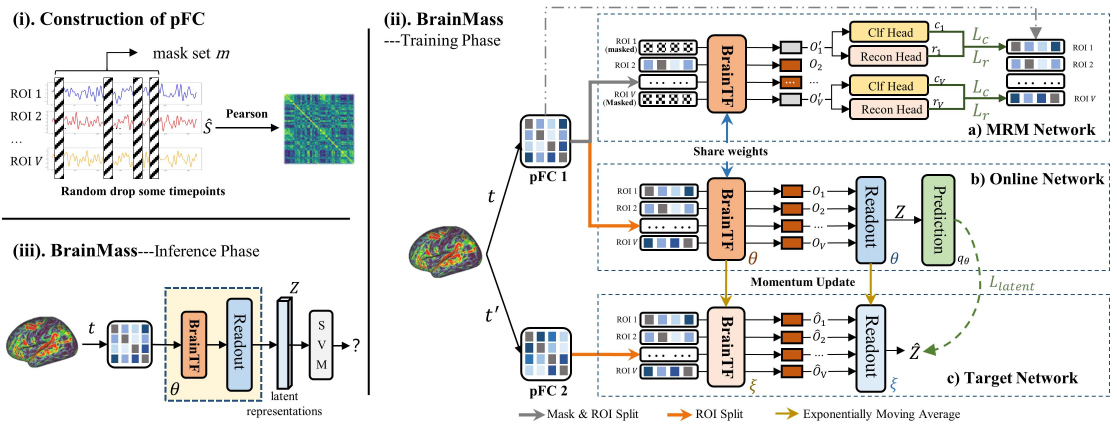}
  \caption{Illustration of (i) the construction of pFC, (ii) the training phase of BrainMass method, including an MRM (an MRM network) and an LRA (an online network and a target network) module, and (iii) the inference phase of BrainMass.}
  \label{model}
\end{figure*}

\section{Method}\label{method}

\subsection{Preliminaries}
The brain functional networks $\mX$ are derived by mapping processed neuroimages onto a template with $V$ Regions of Interest (ROIs). These networks are symmetric positive definite matrices, $\mX \in \R^{V \times V}$. For diagnosis purposes, the goal is to develop a mapping function $f: \mX \rightarrow y$, where $y$ represents the predicted diagnosis phenotype for each subject. 

In this study, we first generate two pFCs for each participant, and feed them into the BrainMass framework for pre-training a brain network Transformer  (BrainTF) encoder. 
During the downstream classification phase, we froze the BrainTF and use it to extract latent representations, $\mZ$, for each participant. 
The learned latent representations are further fed into a Support Vector Machine (SVM) classifier for downstream prediction. This process is shown in Fig. \ref{model}. To note that, in the training phase, the BrainMass 
consists of three components: the MRM network, the online network, and the target network. 
Each network features a BrainTF encoder, sharing the same architectural design. The BrainTFs in the MRM and online networks share the same weights, while the BrainTF in the target network is updated by an exponential moving average based on the online network.





\subsection{Pseudo functional connectivity augmentation}\label{method_pfc}

In this study, we propose to investigate brain network augmentation methods by random removal of certain timepoints within timeseries data. Consider a timeseries matrix $\mS \in \R^{V \times T}$, where $T$ represents the number of time steps. We generate a random dropping vector $\vm \in \R^{M}$, with $M \leq T$, and use this vector to omit selected timepoints from the data, which is shown in Fig. \ref{model} (i). Subsequently, we apply the Pearson correlation to the resulting modified timeseries matrix $\hat{\mS} \in \R^{V \times (T-M)}$ to create a pseudo-functional connectivity (pFC) matrix $\hat{\mX}$.
It is important to note that fMRI data is often collected using a variety of protocols, resulting in differences in lengths of the timeseries data across samples. To address these variations, we investigate the effects of using different percentages of dropping lengths to account for the inherent variability in fMRI data acquisition and enhances the robustness and applicability of the augmentation.

\subsection{Brain network Transformer encoder}\label{method_bntf}
Transformer-based models have led a tremendous success in various downstream tasks across fields including natural language processing, computer vision, and also graph learning. However, the brain networks potentially fall in neither of these classes. The brain networks are symmetric semi-positive defined matrices and densely distributed. Previous studies tackle the brain network as graph data, however, there are still no explicit inter-ROI relationships \cite{kan2022brain,yang2023deep,li2021braingnn}. In this study, we instead tackle the connection profile as a sequence, where each ROI is represented as a sequential step with $V$ features. The input brain network $\mX$ is viewed as a sequence $\{\vx_0,\vx_1,...,\vx_{V-1}\}$, where the $i$-th element is obtained by $\vx_i=\mX_{i,:} \in \R^{V}$. 
In this context, Multi-Head Self-Attention (MHSA) is implemented to relate inter-ROI dependencies and generate more expressive brain features:
\begin{equation}
  \mH^L=\text{MHSA}(\mX) \in \R^{V \times V}
\end{equation}

For each layer $l$, we first calculate the query $\mQ^{l,c}$, key $\mK^{l,c}$, and value $\mV^{l,c}$ for the $c$-th head through linear projection as:
\begin{gather}
  \mQ^{l,c}=\mH^{l-1}\mW^{l,c}_q,\\
  \mK^{l,c}=\mH^{l-1}\mW^{l,c}_k,\\
  \mV^{l,c}=\mH^{l-1}\mW^{l,c}_v
\end{gather}
where $\mH^{l-1}$ is the output of the $l$-th layer, $\mH^0=\mX$, and $W^{l,c}_q, W^{l,c}_k,W^{l,c}_v$ are learnable parameters. $c$ is in the range of $\{1,2,...,C\}$, and $C$ denotes the number of self-attention heads. The output for each head is computed as:
\begin{equation}
  \mH^{l,c}=\text{Softmax}(\frac{\mQ^{l,c}(\mK^{l,c})^T}{\sqrt{d}})\mV^{l,c}
\end{equation}
where $d$ is the first dimension of $\mW^{l,c}$. 
Finally, the output $\mH^l$ is obtained by $\mH^l=(||_{c=1}^C \mH^{l,c})\mW^l_O$, where $||$ is the concatenation operator, and $\mW^l_O$ are learnable model parameters. We implement the Feed Foward Network and layer normalization for mapping the $\mH^l$ into encoder outputs.

\subsection{Masked ROI Modelling}\label{method_mrm}
For the mask modeling, as illustrated in the MRM network of the Figure \ref{model} (ii), each input brain network is treated as a sequence, divided into $V$ ROIs, and randomly assigned a set $\sP$ of $P$ masked ROI position indices. For each patch that needs to be masked, we replace its patch embedding with a learnable mask embedding. Positional embeddings are added to the patch embeddings, and the resulting data is fed into the BrainTF encoder. Classification (Clf) and Reconstruction (Recon) heads are utilized to predict the indices and features of the masked ROIs from the remaining ROI features.

For each masked patch $\vx^{'}_i$, we obtain a corresponding output $\vo^{'}_i$ from the BrainTF encoder. Subsequently, we pass $\vo^{'}_i$ through both a classification head and a reconstruction head to obtain outputs $\vc_i$ and $\vr_i$, respectively. Both the classification and reconstruction heads consist of two-layer MLPs designed to map $\vo^{'}_i$ to the same dimension as $\vx^{'}_i$. The goal is to make $\vr_i$ as close as possible to $\vx^{'}_i$ while ensuring the model can correctly match pairs $(\vx^{'}_i, \vc_i)$.
To achieve this, we employ the InfoNCE loss \cite{oord2018representation} $L_c$ for the classification objective and the mean square error loss $L_r$ for the reconstruction objective:
\begin{equation}
   L_c = -\frac{1}{N}\sum^N_{i=1}\log\left(\frac{\exp(\vc^T_i \vx^{'}_i)}{\sum_{j=1}^N \exp(\vc_i^T \vx^{'}_j)}\right)
\end{equation}

\begin{equation}
   L_r = \frac{1}{N}\sum_{i=1}^N ||\vr_i - \vx^{'}_i||^2
\end{equation}

\begin{table*}[htbp]
  \centering
  \setlength{\tabcolsep}{3.5pt}
  \renewcommand\arraystretch{0.8}
  \scriptsize
  \caption{Demographical information on 30 datasets.}
    \begin{tabular}{cccccccccc}
    \hline
    \hline
    ID    & Type for use & Dataset & Participants & Samples & Age (Mean/Std) & Gender (M/F) & Groups & Pretrain Samples & Downstream Samples \bigstrut\\
    \hline
    1     & \multirow{20}[2]{*}{Type-I} & UKBiobank \cite{sudlow2015uk} & 21240 & 21240 & 54.90/7.49 & 10069/11171 & Multiple & 21240 & - \bigstrut[t]\\
    2     &       & CMI-HBN \cite{alexander2017open}  & 2228  & 4134  & 10.35/4.11 & 1428/800 & Multiple & 4134  & - \\
    3     &       & GSP \cite{holmes2015brain}  & 1570  & 2708  & 21.42/2.89 & 665/905 & -     & 2708  & - \\
    4     &       & CORR \cite{zuo2014open}  & 1515  & 4247  & 25.85/15.38 & 756/759 & -     & 4247  & - \\
    5     &       & NKI-RS \cite{tobe2022longitudinal}  & 1319  & 6863  & 39.00/22.07 & 515/804 & -     & 6863  & - \\
    6     &       & HCP \cite{van2013wu}  & 1206  & 4176  & (20-40) & 550/656 & -     & 4176  & - \\
    7     &       & QTIM \cite{sinclair2015heritability} & 1202  & 1202  & 21.17/4.03 & 470/732 & -     & 1202  & - \\
    8     &       & FCON-1000 & 1070  & 1118  & 28.64/13.49 & 490/580 & -     & 1118  & - \\
    9     &       & SLIM \cite{wei2018structural} & 1008  & 1008  & 20.01/1.20 & 451/557 & -     & 1008  & - \\
    10    &       & CCNP \cite{ccnp} & 878   & 1581  & 11.15/3.04 & 469/409 & -     & 1581  & - \\
    11    &       & CAM-CAN \cite{taylor2017cambridge} & 652   & 652   & 54.85/18.54 & 322/330 & -     & 652   & - \\
    12    &       & CHINA-Project & 608   & 608   & 62.48/9.98 & 308/300 & AD, PD & 608   & - \\
    13    &       & SALD \cite{wei2018structural} & 493   & 493   & 45.15/17.43 & 185/308 & -     & 493   & - \\
    14    &       & INDI-Retro  & 479   & 1494  & 33.19/16.82 & 239/240 & -     & 1474  & - \\
    15    &       & QTAB \cite{strike2023queensland}  & 417   & 1142  & 11.56/1.65 & 215/202 & -     & 1142  & - \\
    16    &       & NAD \cite{spreng2022neurocognitive}  & 300   & 1761  & 40.94/23.08 & 132/168 & -     & 1761  & - \\
    17    &       & ISYB \cite{isyb} & 215   & 215   & 22.60/2.66 & 59/156 & -     & 215   & - \\
    18    &       & CogTrain \cite{kable2017no} & 166   & 210   & 24.52/4.49 & 98/68 & -     & 210   & - \\
    19    &       & Caltech Conte Center \cite{kliemann2022caltech} & 117   & 305   & 28.53/6.41 & 68/49 & -     & 305   & - \\
    20    &       & Synaesthesia \cite{racey2023open} & 126   & 505   & 35.64/13.35 & 28/98 & -     & 505   & - \bigstrut[b]\\
    \hline
    21    & \multirow{6}[2]{*}{\makecell{Type-II\\(Internal)} } & REST-MDD \cite{chen2022direct} & 2379  & 2379  & 36.20/15.11 & 925/1454 & MDD   & 1666  & 1276 MDD, 1104 NC \bigstrut[t]\\
    22    &       & ADHD-200 & 1258  & 1258  & 11.72/3.29 & 765/493 & ADHD  & 882   & 548 ADHD, 710 NC \\
    23    &       & OASIS \cite{lamontagne2019oasis} & 1185  & 4516  & 70.06/9.43 & 446/739 & DM & 2222  & 309 DM, 300 NC \\
    24    &       & ADNI \cite{jack2008alzheimer}  & 1171  & 3050  & 71.25/7.09 & 560/611 & MCI, AD & 2529  & 151 MCI, 149 AD, 142 NC \\
    25    &       & ABIDE-I & 1114  & 1114  & 16.86/8.06 & 957/157 & ASD   & 779   & 528 ASD, 556 NC \\
    26    &       & PPMI \cite{marek2011parkinson} & 941   & 973   & 65.15/8.51 & 548/393 & PD    & 864   & 70 PD, 70 proPD, 64 NC \bigstrut[b]\\
    \hline
    27    & \multirow{4}[2]{*}{\makecell{Type-III\\(External)} } & ABIDE-II & 1236  & 1236  & 14.66/9.12 & 895/341 & ASD   & 0     & 559 ASD, 677 NC \bigstrut[t]\\
    28    &       & LA5c \cite{poldrack2016phenome} & 192   & 192   & 34.04/9.32 & 118/74 & Multiple & 0     & 43 ADHD, 49 BP, 50 SCZ, 50 NC \\
    29    &       & Xuanwu \cite{yang2021alteration} & 213   & 213   & 61.72/9.61 & 107/106 & PD, iRBD & 0     & 90 PD, 53 iRBD, 70 NC \\
    30    &       & SchizoConnect \cite{wang2016schizconnect} & 188   & 188   & 37.90/12.76 & 143/45 & SCZ & 0     & 97 SCZ, 91 NC \bigstrut[b]\\
    \hline
          &       & Total & 46686 & 70781 &       & 23018/23748 &       & 64584 &  \bigstrut\\
    \hline
    \hline
    \end{tabular}%
  \label{data}%
  \\
  \smallskip
  \begin{flushleft}
  \scriptsize{
    NC: Normal Control, ASD: Autism Spectrum Disorder, ADHD: Attention Deficit Hyperactivity Disorder, DM: Dementia, PD: Parkinson's Disease, proPD: Prodromal Parkinson's Disease, BP: Bipolar Disorder, iRBD: Idiopathic Rapid Eye Movement
  Sleep Behavior Disorder, MDD: Major Depression Disorder, MCI: Mild Cognitive Impairment, AD: Alzheimer's Disease, SCZ: Schizophrenia}
   \\
  \tiny{
  ADHD-200: \url{https://fcon_1000.projects.nitrc.org/indi/adhd200/},
  INDI-Retro: \url{https://fcon_1000.projects.nitrc.org/indi/IndiRetro.html},\\
  FCON-1000: \url{https://fcon_1000.projects.nitrc.org/fcpClassic/FcpTable.html}, ABIDE: \url{https://fcon_1000.projects.nitrc.org/indi/abide}.
  
  }
  \end{flushleft}
\end{table*}%

\subsection{Latent representation alignment}\label{method_lra}



BrainMass leverages the latent representation alignment on two pFCs to achieve similar latent representations. Following previous works \cite{grill2020bootstrap,chen2020improved}, we employ a dual branch including the online network and the target network. The online network encodes brain networks using an $L$-layer Multi-head Self-Attention (MHSA) Transformer network, resulting in nonlinear mappings as $\mX \rightarrow \mO \in \R^{V \times V}$. A readout function subsequently transforms the encoded features $\mO$ into embeddings $\mZ \in \R^{D \times V}$. Similarly, the target network generates embeddings $\hat{\mZ}$ through the same process. The target network provides regression targets for training the online network. To prevent model collapse, a predictor, $q_{\theta}$, is used to maintain asymmetry between the online and target networks. A prediction Multi-Layer Perceptron (MLP) is employed to learn the mapping from the outputs $\mZ$ of the online network to predict the outputs $\hat{\mZ}$ of the target network. Parameters of the target network $\xi$ are updated using an exponentially weighted moving average based on the online parameters $\theta$:
\begin{equation}
  \xi \leftarrow \tau \xi + (1-\tau)\theta
\end{equation}
where $\tau$ is a target decay rate $\tau \in [0,1]$.
The optimization is performed using the mean squared error between the normalized predictions and the target projections:

\begin{equation}
   L_{latent}=2 - 2 \frac{<q_{\theta}(\mZ), \hat{\mZ}>}{||q_{\theta}(\mZ)||_2 \cdot ||\hat{\mZ}||_2}
\end{equation}

\textbf{Readout function.} After obtaining the non-linear features from the Transformer encoders, we are left with high-dimensional data, which pose challenges for downstream classification tasks, especially given the limited fMRI data samples available. In this study, we employ a readout function to transform the output features $\mO$ into dimension-reduced embeddings. To achieve this, we aggregate the output features for each ROI into a set of $D$ features. These features are then concatenated to form the final feature representation $\mO \in \R^{D \times V}$. In this study, we set $D=8$ and finally obtain $8 \times V$ representations for each brain network.

\subsection{Optimization}
To finalize the objective function for self-supervised pre-training, we employ a weighted summation that includes balancing parameters $\lambda_c$ and $\lambda_r$ as:

\begin{equation}\label{loss}
   L = L_{latent} + \lambda_c L_c + \lambda_r L_r
\end{equation}
This regularization enables the brainTF encoder to learn intra-network dependencies across various brain connectome patterns, enhancing its effectiveness for downstream tasks.


\section{Experiments}\label{exp}

\textbf{Datasets.}
We built a large cohort for both pretraining and evaluation, with detailed demographic information presented in Table \ref{data}. This cohort includes participants from multiple centers, races, and countries throughout the world. To the best of our knowledge, this is the largest dataset for brain network analysis, featuring a diverse group of participants comprising a range of diagnosis types, including Normal Control (NC) subjects and individuals with various neurodevelopmental and neurodegenerative disorders. These disorders encompass Major Depression Disorder (MDD), Mild Cognitive Impairment (MCI), Alzheimer's Disease (AD), Autism Spectrum Disorder (ASD), Attention Deficit Hyperactivity Disorder (ADHD), Dementia (DM), Parkinson's Disease (PD), Prodromal PD (proPD), Bipolar Disorder (BP), and Idiopathic Rapid Eye Movement Sleep Behavior Disorder (iRBD).


The datasets are categorized into three types based on their utilization:
(1) Type-I Datasets (Pre-training): 20 datasets are reserved for pre-training purposes. 
(2) Type-II Datasets (Pre-training and internal evaluation): Six datasets are used both for pre-training and evaluation. From these, 70\% of samples are randomly selected and set fixed use for pre-training and downstream training. 
(3) Type-III Datasets (External evaluation): Four datasets are designated as external datasets. These are crucial for assessing  the model's generalization and adaptability capabilities.
A total of 64,584 samples are incorporated into the pre-training phase. 
Ten datasets are specifically employed to evaluate the diagnosis of brain diseases. Notably, for fair comparison, single scan of fMRI for each subject was collected in the evaluation, and the duplicated scans are excluded for both pretraining and downstream evaluations. 

\begin{table*}[htbp]
  \centering
  \renewcommand\arraystretch{0.95}
  \scriptsize 
\setlength{\tabcolsep}{2pt}
  \caption{Classification results of different approaches on 8 tasks of 6 internal datasets  in terms of accuracy (Acc), sensitivity (Sen), and specificity (Spe). SSL indicates the model is pretrained by self-supervised learning.}

\begin{tabular}{c|c|cccccccccccccccc}
  \hline
  \hline
  Dataset & \multirow{3}[6]{*}{SSL} &       & \multicolumn{3}{c}{ABIDE-I} &       & \multicolumn{3}{c}{ADHD-200} &       & \multicolumn{3}{c}{REST-MDD*} &       & \multicolumn{3}{c}{OASIS} \bigstrut\\
  \cline{4-6}\cline{8-10}\cline{12-14}\cline{16-18}Task  &       &       & \multicolumn{3}{c}{NC vs. ASD} &       & \multicolumn{3}{c}{NC vs. ADHD} &       & \multicolumn{3}{c}{NC vs. MDD} &       & \multicolumn{3}{c}{NC vs. DM} \bigstrut\\
  \cline{4-6}\cline{8-10}\cline{12-14}\cline{16-18}Metric &       &       & ACC  $\uparrow$ & SEN $\uparrow$  & SPE $\uparrow$  &       & ACC $\uparrow$  & SEN $\uparrow$  & SPE $\uparrow$  &       & ACC $\uparrow$  & SEN $\uparrow$  & SPE $\uparrow$  &       & ACC $\uparrow$  & SEN $\uparrow$  & SPE $\uparrow$\bigstrut\\
  \hline
  BrainNetCNN &       &       & 68.14-2.04 & 67.56-7.02 & 69.74-4.41 &       & 61.62-1.81 & 63.18-8.39 & 61.82-1.46 &       & 62.55-1.81 & \underline{64.41-8.39} & 59.93-1.46 &       & 68.24-3.66 & 67.42-4.21 & 69.53-3.22 \bigstrut[t]\\
  DHGNN &       &       & 64.31-1.52 & 63.81-5.03 & 64.97-2.62 &       & 59.84-2.04 & 53.52-2.65 & 61.72-2.51 &       & 59.24-2.04 & 61.40-2.65 & 56.51-2.51 &       & 66.71-5.61 & 66.39-4.90 & 67.29-6.88 \\
  BrainGNN &       &       & 69.60-2.24 & 61.47-3.59 & 71.46-2.57 &       & 61.02-2.59 & 54.60-4.05 & 64.08-2.85 &       & 61.40-2.59 & 61.37-4.05 & 55.86-2.85 &       & 68.24-2.27 & 61.74-9.29 & \underline{74.89-7.31} \\
  PopGCN &       &       & 69.76-1.40 & 67.61-3.12 & 71.72-1.74 &       & 62.20-1.36 & 56.69-2.17 & \textbf{66.40-2.98} &       & 61.20-1.36 & 64.06-2.17 & 57.23-2.98 &       & 65.93-3.64 & 66.00-4.35 & 65.82-3.24 \\
  vanillaTF    &       &       & 68.98-1.13 & 65.01-5.19 & \textbf{72.48-4.03} &       & 61.62-1.14 & 63.18-5.20 & 61.82-1.74 &       & 62.49-1.14 & \textbf{64.61-5.20} & 60.65-1.74 &       & 68.57-2.26 & 68.38-3.80 & 69.74-3.31 \\
  BrainNetTF &       &       & \underline{71.02-1.16} & \underline{73.27-5.62} & 71.18-4.38 &       & 62.75-1.29 & \underline{63.61-5.25} & 62.85-2.25 &       & \underline{63.50-1.14} & 64.05-5.20 & 61.56-1.74 &       & \underline{72.53-2.41} & \underline{74.55-5.58} & 71.41-2.26 \\
  MoCo  & \checkmark &       & 68.68-2.50 & 66.01-2.97 & 70.85-3.45 &       & 58.95-2.77 & 51.02-6.25 & 62.66-2.89 &       & 62.34-1.55 & 61.64-1.12 & \underline{63.83-2.76} &       & 71.32-3.88 & 69.48-3.71 & 74.18-5.57 \\
  BYOL  & \checkmark &       & 68.98-1.49 & 67.88-3.75 & 70.00-2.82 &       & 59.46-3.55 & 51.98-8.60 & 63.12-2.28 &       & 62.80-1.16 & 62.43-0.82 & 63.61-2.15 &       & 68.79-3.97 & 70.17-4.66 & 67.70-3.79 \\
  BrainNPT & \checkmark &       & 63.83-2.84 & 61.51-4.32 & 65.51-2.99 &       & 58.27-2.52 & 55.77-9.39 & 58.68-2.24 &       & 57.84-1.31 & 58.84-0.97 & 56.01-2.03 &       & 64.51-2.74 & 65.02-3.58 & 65.01-4.15 \\
  BrainGSLs & \checkmark &       & 70.98-3.68 & 70.62-4.10 & 71.71-4.94 &       & \underline{62.32-2.96} & 61.48-5.28 & \underline{66.25-5.42} &       & 59.87-2.67 & 59.11-3.10 & 62.22-1.23 &       & 59.87-2.67 & 59.11-3.10 & 62.22-1.23 \\
  BrainMass & \checkmark &       & \textbf{72.75-2.85} & \textbf{74.26-3.88} & \underline{71.89-5.13} &       & \textbf{64.65-1.66} & \textbf{64.36-6.16} & 64.85-1.38 &       & \textbf{66.42-1.16} & \textbf{64.61-1.24} & \textbf{67.61-1.41} &       & \textbf{76.48-2.26} & \textbf{75.14-2.08} & \textbf{78.25-3.53} \bigstrut[b]\\
  \hline
  \hline
  Dataset & \multirow{3}[6]{*}{SSL} &       & \multicolumn{3}{c}{ADNI} &       & \multicolumn{3}{c}{ADNI} &       & \multicolumn{3}{c}{PPMI} &       & \multicolumn{3}{c}{PPMI} \bigstrut\\
  \cline{4-6}\cline{8-10}\cline{12-14}\cline{16-18}Task  &       &       & \multicolumn{3}{c}{NC vs. MCI} &       & \multicolumn{3}{c}{NC vs. AD} &       & \multicolumn{3}{c}{NC vs. proPD} &       & \multicolumn{3}{c}{NC vs. PD} \bigstrut\\
  \cline{4-6}\cline{8-10}\cline{12-14}\cline{16-18}Metric &       &       & ACC $\uparrow$  & SEN $\uparrow$  & SPE $\uparrow$  &       & ACC $\uparrow$  & SEN $\uparrow$  & SPE $\uparrow$  &       & ACC $\uparrow$  & SEN $\uparrow$  & SPE $\uparrow$  &       & ACC $\uparrow$  & SEN $\uparrow$  & SPE $\uparrow$ \bigstrut\\
  \hline
  BrainNetCNN &       &       & 56.51-3.90 & 59.12-6.76 & 55.61-5.01 &       & 67.21-4.34 & 69.76-7.54 & 67.69-5.63 &       & 69.44-6.69 & 62.22-6.78 & 81.69-8.80 &       & 73.00-6.40 & 70.24-6.62 & 77.56-7.02 \bigstrut[t]\\
  DHGNN &       &       & 52.09-4.90 & 53.02-4.07 & 48.43-1.72 &       & 61.40-4.19 & 63.45-7.01 & 62.51-5.97 &       & 71.67-3.89 & 63.95-4.66 & 86.18-8.08 &       & 70.50-5.68 & 68.38-4.99 & 77.44-10.24 \\
  BrainGNN &       &       & 61.16-5.61 & 55.46-9.04 & \textbf{67.14-10.31} &       & 71.63-3.42 & 74.55-8.18 & 68.57-10.09 &       & 67.22-5.58 & 63.89-8.29 & 84.89-3.46 &       & 76.50-4.50 & \underline{79.00-5.39} & 64.00-11.14 \\
  PopGCN &       &       & 61.43-7.00 & 61.61-7.17 & 60.84-7.57 &       & 66.16-5.76 & 65.56-6.60 & 66.15-6.04 &       & 67.78-5.24 & 62.98-5.90 & 73.44-7.81 &       & 67.50-6.80 & 66.97-7.77 & 67.48-6.91 \\
  vanillaTF    &       &       & 57.91-5.45 & 60.03-6.21 & 57.54-7.09 &       & 72.56-4.51 & 69.84-6.14 & 78.48-5.90 &       & 72.22-3.51 & 64.41-3.73 & 84.60-6.67 &       & 74.00-7.68 & 70.87-7.59 & 81.11-11.42 \\
  BrainNetTF &       &       & \underline{62.10-3.46} & \underline{64.53-4.70} & 61.44-4.67 &       & 75.81-3.78 & 81.05-9.44 & 73.77-4.68 &       & 72.22-4.97 & 65.76-7.64 & 83.00-6.81 &       & 77.00-9.27 & 73.95-9.64 & 82.53-9.43 \\
  MoCo  & \checkmark &       & 60.70-5.14 & 59.46-4.16 & 63.19-7.16 &       & 74.42-5.79 & 75.29-6.03 & 74.26-7.19 &       & 77.22-7.64 & 73.20-9.07 & 83.53-9.11 &       & \underline{83.50-4.50} & 70.64-5.14 & \textbf{88.68-8.36} \\
  BYOL  & \checkmark &       & 56.74-3.16 & 57.32-2.87 & 56.21-3.83 &       & 74.93-5.91 & 74.63-5.30 & 75.80-7.46 &       & \underline{77.78-8.96} & 72.52-9.38 & 86.88-9.65 &       & 64.00-7.00 & 61.22-6.01 & 69.67-9.76 \\
  BrainNPT & \checkmark &       & 55.58-4.09 & 55.25-3.44 & 56.21-5.50 &       & 66.98-3.72 & 68.84-2.46 & 59.72-8.60 &       & 72.78-8.33 & 69.54-9.58 & 76.81-11.18 &       & 61.50-4.50 & 61.00-3.00 & 65.50-2.55 \\
  BrainGSLs & \checkmark &       & 57.33-7.10 & 60.23-7.86 & 60.27-7.08 &       & \underline{76.64-3.28} & \underline{81.12-8.78} & \textbf{82.88-5.36} &       & 76.35-7.86 & \underline{73.52-9.03} & \underline{87.46-4.61} &       & 77.53-8.56 & 78.38-9.18 & 79.97-8.68 \\
  BrainMass & \checkmark &       & \textbf{68.37-3.48} & \textbf{70.51-5.07} & \underline{66.61-2.71} &       & \textbf{83.72-2.55} & \textbf{85.94-5.86} & \underline{82.29-2.79} &       & \textbf{80.00-4.44} & \textbf{74.70-9.29} & \textbf{88.18-8.66} &       & \textbf{84.50-6.87} & \textbf{82.69-7.63} & \underline{87.31-7.57} \bigstrut[b]\\
  \hline
  \hline
  \end{tabular}%

  \label{res}%
\end{table*}%

\textbf{Data Preprocessing.}
All the fMRI images were pre-processed by reference to the Configurable Pipeline for the Analysis of Connectomes (C-PAC) pipeline , including skull striping, slice timing correction, motion correction, global mean intensity normalization, nuisance signal regression with 24 motion parameters, and band-pass filtering (0.01-0.08Hz). The functional images were finally registered into standard anatomical space (MNI152). The mean time series for a set of regions were computed and normalized into zero mean and unit variance. Pearson Coefficient Correlation was applied to measure functional connectivity.
In this study, the pre-processed fMRI images were mapped by the brain template for parcellations by the Schaefer atlas \cite{schaefer2018local} into 100 ROIs.

\textbf{Implementation details.} 
For pretraining, we utilized the Adam optimizer with an initial learning rate of $3 \times 10^{-5}$ and a weight decay of $5 \times 10^{-5}$. The learning rate underwent a linear increase to $3 \times 10^{-4}$ within 10 warmup epochs. The batch size was set fixed as 256. The BrainMass model underwent training for 2000 epochs, and we saved the models with the lowest training loss for subsequent classification tasks. The decay rate $\tau$ for target network update is set as $0.996$. Our experiments were conducted on a platform equipped with 64 NVIDIA Tesla V100 GPUs, with 8 GPUs allocated for each training. It takes around 150 hours for each pretraining. The implemented Transformer encoder is configured with 32 layers, and 20 heads for MHSA. The hidden feature dim is 4096 for FFN. The total parameter size is $67.0 \text{M}$.
For the optimization, we set $\lambda_c$ and $\lambda_r$ to 0.1 and 5 in Eq. \ref{loss}. In the downstream tasks, the latent representations were input into an SVM classifier for prediction. To facilitate a more robust comparison on smaller-sized datasets, we repeated the downstream tasks 10 times by randomly sampling the validation and test sets.

\textbf{Metrics.} We assess the performance of diagnosis classification using accuracy (ACC), sensitivity (SEN), and specificity (SPE) as our key metrics. 
We employ a rigorous stratified sampling strategy that considers collection sites during the training (70\%)-validation (15\%)-testing (15\%) split, ensuring fair comparisons \cite{kan2022brain}. 

\begin{table*}[htbp]
  \centering
  \scriptsize
  \renewcommand\arraystretch{0.9}
  \caption{Ablation studies on the elements of BrainMass with the accuracy (\%) performance on eight internal tasks.}

\begin{tabular}{ccc|cccccccccccccc}
  \hline
  \hline
  \multirow{2}[3]{*}{Latent} & \multicolumn{2}{c|}{MRM} &       & ABIDE-I &       & ADHD-200  &       & REST-MDD &       & OASIS   &       & \multicolumn{2}{c}{ADNI} &       & \multicolumn{2}{c}{PPMI} \bigstrut[b]\\
  \cline{2-3}\cline{5-5}\cline{7-7}\cline{9-9}\cline{11-11}\cline{13-14}\cline{16-17}      & $L_c$ & $L_r$ &       & ASD   &       & ADHD  &       & MDD   &       & DM    &       & MCI   & AD    &       & proPD & PD \bigstrut\\
  \hline
  \checkmark &       &       &       & 63.83-2.84 &       & 58.27-2.52 &       & 62.80-1.16 &       & 64.51-2.74 &       & 53.26-4.47 & 70.93-5.91 &       & 77.78-8.96 & 64.00-7.00 \bigstrut[t]\\
        &       & \checkmark &       & 65.39-3.01 &       & 60.76-1.90 &       & 64.65-2.67 &       & 69.01-4.28 &       & 55.12-4.54 & 66.05-5.32 &       & 77.78-3.51 & 74.00-7.68 \\
  \checkmark & \checkmark &       &       & 67.13-2.48 &       & 60.16-1.43 &       & 62.46-2.44 &       & 67.98-2.66 &       & 61.40-3.15 & 72.09-3.75 &       & 74.44-6.67 & \textbf{84.50-6.50} \\
  \checkmark &       & \checkmark &       & 71.62-2.84 &       & 61.62-2.84 &       & 66.00-1.31 &       & 71.98-2.96 &       & 61.86-4.67 & 79.07-2.55 &       & \textbf{80.00-3.68} & \textbf{84.50-6.50} \\
  \checkmark & \checkmark & \checkmark &       & \textbf{72.75-2.85} &       & \textbf{64.65-1.66} &       & \textbf{66.42-1.16} &       & \textbf{76.48-2.26} &       & \textbf{68.37-3.48} & \textbf{83.72-2.55} &       & \textbf{80.00-4.44} & \textbf{84.50-6.87} \bigstrut[b]\\
  \hline
  \hline
\end{tabular}%

  \label{ablation}%
\end{table*}%

\section{Results}\label{res_sec}

\subsection{Brain disorder diagnosis performance}

For comparison, two categories of baseline models are included: those with SSL and those without SSL.
The baseline models without SSL include BrainNetCNN \cite{kawahara2017brainnetcnn}, DHGNN \cite{jiang2019dynamic}, BrainGNN \cite{li2021braingnn}, Semi-GCN \cite{parisot2018disease}, vanilla-Transformer (vanillaTF), and BrainNetTransformer (BrainNetTF) \cite{kan2022brain}. 
For SSL comparisons, powerful SSL frameworks like BYOL \cite{grill2020bootstrap} and MOCO \cite{he2020momentum} are included. Furthermore, we considered two existing works: BrainNPT \cite{hu2023brainnpt} and BrainGSLs \cite{wen2023graph}.

Table \ref{res} presents the results from 8 tasks across 6 internal datasets, with the highest performance marked in bold and the second-best underlined. Notably, for the REST-MDD dataset, as raw images were not available, we additionally trained a model by mapping fMRIs with the AAL atlas into 116 ROIs. From these results, we observe the following key points:
1) Among these models, CNN, GNN, and Transformer architectures demonstrate similar performance across all tasks. Despite their increased computational complexity, Transformer models offer limited performance enhancement when handling fMRI data with limited sample sizes. However, BrainNetTF significantly boosts diagnostic performance, corroborating findings from previous studies \cite{kan2022brain}.
2) Contrary to expectations, most SSL approaches did not markedly enhance performance in comparison to BrainNetTF. In fact, some SSL methods even underperformed relative to baseline models without SSL.
3) Our BrainMass model consistently outperforms these methods across all 8 tasks, with accuracy improvements of 1.73\%, 2.33\%, 2.92\%, 3.95\%, 6.27\%, 7.08\%, 2.22\%, and 1.00\% for distinguishing ASD, ADHD, MDD, DM, MCI, AD, proPD, and PD, respectively. This highlights the significant benefits of large-scale SSL representations and underscores the effectiveness of our BrainMass framework.

Additionally, while previous studies \cite{parisot2018disease,yang2023creg} demonstrate higher accuracy in distinguishing AD/MCI from NC, our results indicate lower overall accuracy. On one hand, our study utilizes a single scan per participant, unlike others that incorporate multiple scans to expand the dataset. Repeated scans might exhibit similar functional features, boosting accuracy. Compared with them, in our setting, multimodal approaches could only enhance the accuracy to 88.6\% in our previous studies \cite{yang2023mapping}. On the other hand, we adopt a rigorous model selection strategy by choosing the optimal model based on the validation set, aligning with \cite{kan2022brain,parisot2018disease}, which might affect generalized performance due to the validation-test gap. 

\subsection{Sensitive analysis and ablation studies}
\textbf{Ablation studies.}
We undertook evaluations focusing on the distinct components of Eq. \ref{loss}. These components include latent representation learning ($L_{latent}$), as well as MRM with classification heads ($L_c$) and reconstruction heads ($L_r$) for distinguishing masked ROI indices and features. The results corresponding to these evaluations are compiled in Table \ref{ablation}.
We can find that incorporating the $L_r$ term significantly enhances performance across all tasks. While the ROI meta-label prediction term ($L_c$) has a comparatively modest impact on its own, its integration with $L_{latent} + L_r$ yields further improvements in model performance. This enhancement can be attributed to the model's increased proficiency in capturing dependencies among brain regions and in learning intra-network representations. The masked ROI distinguishing module plays a pivotal role in this process, facilitating a more nuanced and effective learning paradigm. When we finally combined all these elements, the performances are further improved in most cases.

\begin{figure}
  \centering
  \includegraphics[width=1.0\linewidth]{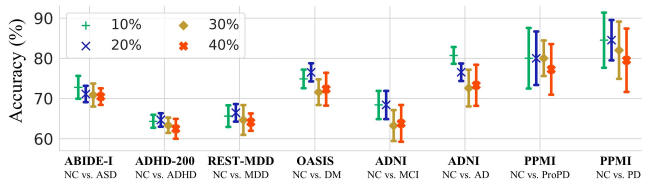}
  \caption{The effect on the dropping rate on eight internal tasks.}
  \label{hyper}
\end{figure}
\begin{figure}
  \centering
  \includegraphics[width=1.0\linewidth]{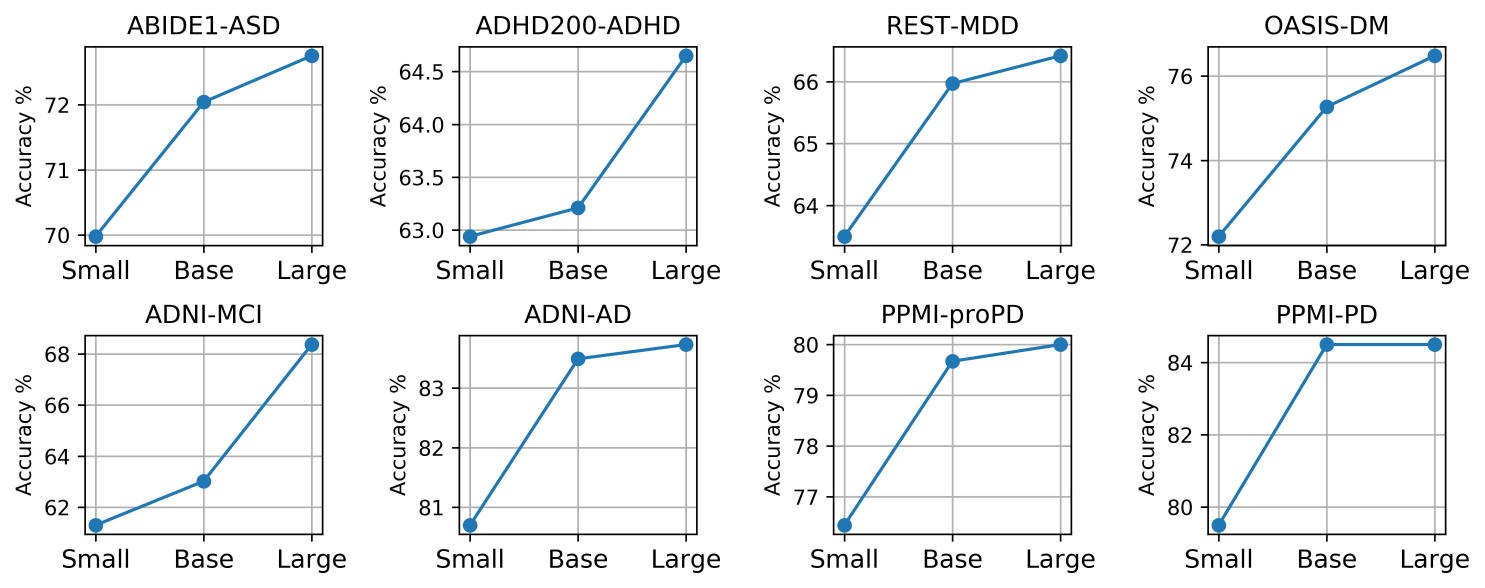}
  \caption{The effect on the model size.}
  \label{model_size}
\end{figure}

\textbf{Drop ratio analysis.}
In this study, we introduce the approach of constructing pFCs to generate millions of brain networks, with a specific focus on mitigating temporal dynamics. A critical aspect of this process is the dropping rate, which we identify as a key hyperparameter. To explore its impact, we analyzed the accuracy performance of our model across various dropping rates, ranging from 10\% to 40\%, as depicted in Figure \ref{hyper} B).
Our observations reveal an initial improvement in performance as the drop ratio increases, followed by a decline once a certain threshold is exceeded. We found that, across all tasks, the optimal dropping rate falls within the range of 10\% to 20\%. Notably, when the dropping ratio surpasses 40\%, there is a consistent deterioration in performance across all 8 tasks.
This pattern highlights the delicate balance required in selecting the appropriate dropping rate for optimal model performance. It suggests that while some degree of dropping can enhance model efficacy by reducing temporal noise, excessive dropping may lead to the loss of critical temporal information, adversely affecting the model's ability to accurately analyze brain networks.


\textbf{Model Size analysis.} 
Figure \ref{model_size} presents the accuracy of models of varying sizes, categorized as small, base, and large. These models are equipped with 8, 16, and 32 Transformer layers, 5, 10, and 20 attention heads, and 1024, 2048, and 4096 FFN features, respectively. Their total parameters amount to 14.4 M, 25.4 M, and 67.0 M, respectively. The results show that the large model configuration achieves superior performance in all tested scenarios. Furthermore, there is a noticeable trend of increasing accuracy as the model becomes larger. The base model demonstrates significant performance improvements in six tasks compared with the small model, except on the ADHD and MCI classification tasks. The large model not only enhances performance slightly across these six tasks but also shows significant improvements in ADHD and MCI classification tasks. Consequently, large models trained with more compute and parameters, exhibit potential emergent abilities with substantial performance increases.

\begin{figure}
  \centering
  \includegraphics[width=0.8\linewidth]{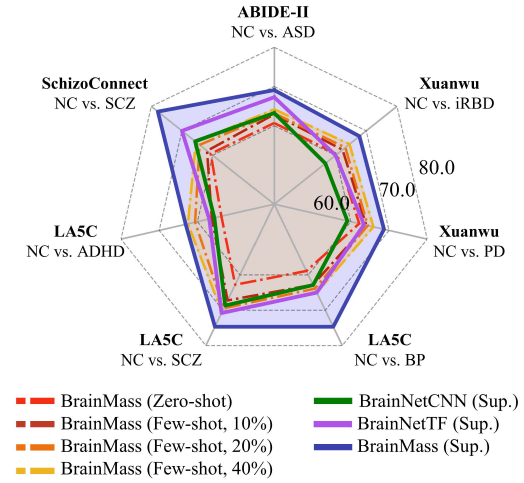}
  \caption{The accuracy performances on seven external tasks.}
  \label{rador}
\end{figure}
\begin{figure*}
  \centering
  \includegraphics[width=0.9\linewidth]{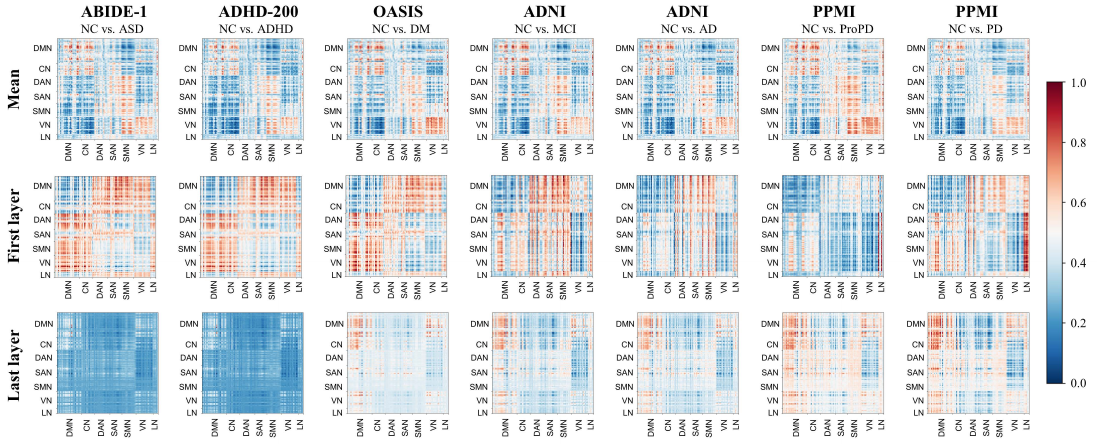}
  \caption{Heatmaps of the Transformer encoder attention maps on 7 tasks, including the averaged attention maps (the first row), those of the first layer (the second row), and the last layer (the third row). The values in heatmaps are normalized into 0 to 1.}
  \label{explain}
\end{figure*}

\begin{figure*}
  \centering
  \includegraphics[width=0.85\linewidth]{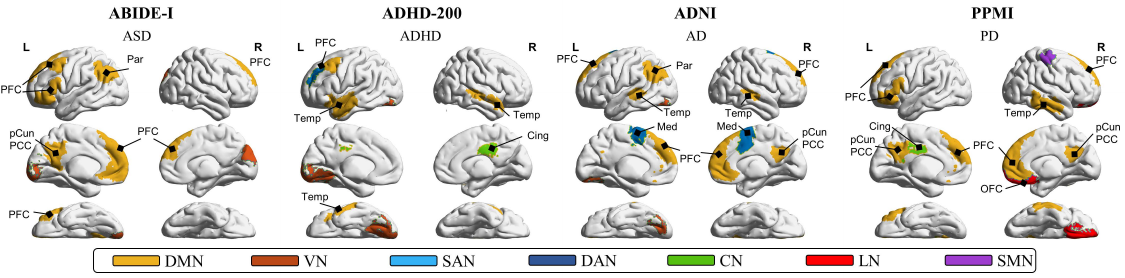}
  \caption{Visualization on the top 10 regions. The key regions are colored with the corresponding sub-network. Temp:
  the temporal. Par: the parietal. Cing: the cingulate. Med: the medial. PFC: the prefrontal cortex. pCun: the precuneus. PCC: the posterior cingulate cortex. OFC: the orbital frontal cortex.}
  \label{bio}
\end{figure*}
\subsection{Generalizability and few/zero-shot evaluation}\label{gen_eva}
We extended our evaluation to external datasets, benchmarking our BrainMass against the baseline BrainNetCNN and the state-of-the-art BrainNetTF. As shown in Figure \ref{rador}, we analyzed the accuracy scores across seven tasks, involving distinguishing NC from ASD, SCZ, ADHD, BP, PD, and iRBD on ABIDE-II, SchizoConnect, LA5C, and Xuanwu datasets. Our BrainMass consistently outperformed both BrainNetCNN and BrainNetTF in these tasks, achieving improvements of 1.81\%, 7.78\%, 4.22\%, 3.57\%, 3.87\%, 6.15\%, and 7.86\% over BrainNetTF, respectively.

Furthermore, our diagnostic task involved distinguishing disorders from NC. To this end, we further studied whether this could be generalized to other diseases. We developed eight classifiers using internal datasets and applied ensemble learning for few-shot and zero-shot learning on external datasets. For zero-shot inference, ensemble weights were calculated by averaging the prediction probabilities of eight classifiers. For few-shot learning, weighted summation was applied based on the probabilities of available samples. The corresponding performances are shown in Fig. \ref{rador} with light red indicating zero-shot, and dark red, orange, and yellow for few-shot with 10\%, 20\%, 40\% samples. From the resutls, we can see that zero-shot inference sometimes equals or exceeds the supervised baseline BrainNetCNN. Moreover, with 20\% annotated samples, BrainMass even surpasses BrainNetTF in differentiating ADHD and iRBD. This highlights BrainMass's remarkable capability in generalizing to various diseases, showcasing its potential for clinical applications with limited annotated samples.
\subsection{Biological explanation}

We present the attention maps across seven internal tasks in Figure \ref{explain}, with the averaged heatmaps, the heatmaps of the first layer, and those of the last layer. Using the Schaefer atlas, the brain network is divided into seven sub-networks: the default mode network (DMN), the visual network (VN), the salience ventral attention network (SAN), the dorsal attention network (DAN), the control network (CN), the somatomotor network (SMN), and the limbic network (LN). 
From the first row, we can see that the BrainTF encoder generally obtains similar averaged heatmaps for different brain disorders. This insight is crucial for zero-shot/few-shot learning, as it provides explainable evidence on our model's ability to differentiate disorders from the normal. 

In addition, across all diseases, we find that the shallower layers focus more on the interactions of DMN and CN with other networks, while this trend shifts in deeper layers. Inter-network communication is significant since the brain is not made up of isolated networks and many tasks require information passing and neuron firing through multiple networks \cite{mahmood2022through}. This communication involves a complex balance among networks with profound implications for understanding human behavior in health and disease \cite{mitra2018principles}. Our model exhibits this characteristic in low-level information processing within shallow layers, especially in interactions between the DMN and other networks, which are critical for cognitive control tasks like attention, memory, and execution. Conversely, at higher levels of information processing within deeper layers, the models discern more disease-specific patterns. 
We suspect that this process closely mirrors the mechanisms of the human brain.
For a cognition task, the initial step involves interactions between different subnetworks, which are crucial for determining how information is processed and how the task is initiated. 
The focus then shifts to specific subnetworks that handle the intrinsic workload of the task. This aligns with research suggesting that cognitive processing involves dynamic and complex interplay among various brain regions, each contributing uniquely to the cognitive function \cite{luppi2022synergistic,ito2017cognitive}. 

In the last layer (the third row), we can see that the disease patterns bifurcate into two categories visually: neurodevelopmental (ASD and ADHD) and neurodegenerative diseases, each with similar patterns within their group. This observation demonstrates that our BrainMass has the abilities to interpret the differences between various diseases. 
To this end, we present the top 10 brain regions in Fig. \ref{bio}, using multivariate analysis \cite{yang2021alteration,shehzad2014multivariate}. 
Consistently found across all diseases is the DMN, emerging as a critical network. It is associated with task-irrelevant mental processes, emotional and self-referential cognitive control, and memory encoding \cite{wei2015association}. 
DMN alterations might contribute to attention lapses and memory deficits observed in AD, PD, ASD, and ADHD. 
Additionally, other crucial biomarker regions, such as the SMN in PD progression and the LN in AD progression, are also identified, consistent with previous studies \cite{caspers2021within,qi2019altered}. 
Overall, these findings suggest that our BrainMass allows for the meaningful interpretation of key biomarkers.

\section{Conclusion}\label{con}
In this study, we developed the first foundation model, BrainMass, specifically designed for brain network analysis. BrainMass leverages the MRM and LRA modules to pre-train the Transformer encoder, focusing on intra-network dependencies and bootstrapped regularized latent representations. This framework aims to foster generalizable and homogeneous representations, thereby enabling highly effective diagnostic performances. Through extensive experimentation involving eight internal and seven external tasks, we validated the efficacy of our framework.
Visualizations on the attention maps and multivariate analysis on the latent representations demonstrate the model's potential emergence ability to discriminate the abnormal from the normal. This highlights its potential for clinical application with robust zero-shot and few-shot learning abilities.
Our study provides new insights into the application of large-scale self-supervised learning in the realm of brain functional network analysis. 

\bibliographystyle{IEEEtran}
\bibliography{bib}

\end{document}

%% file: math_commands.tex
\usepackage{amsmath,amsfonts,bm}

\def\eqref#1{equation~\ref{#1}}

\def\1{\bm{1}}

\def\vc{{\bm{c}}}

\def\vm{{\bm{m}}}

\def\vo{{\bm{o}}}

\def\vr{{\bm{r}}}

\def\vx{{\bm{x}}}

\def\mH{{\bm{H}}}

\def\mK{{\bm{K}}}

\def\mO{{\bm{O}}}

\def\mQ{{\bm{Q}}}

\def\mS{{\bm{S}}}

\def\mV{{\bm{V}}}
\def\mW{{\bm{W}}}
\def\mX{{\bm{X}}}

\def\mZ{{\bm{Z}}}

\DeclareMathAlphabet{\mathsfit}{\encodingdefault}{\sfdefault}{m}{sl}
\SetMathAlphabet{\mathsfit}{bold}{\encodingdefault}{\sfdefault}{bx}{n}

\def\sP{{\mathbb{P}}}

\newcommand{\R}{\mathbb{R}}

